# Ultrafast All-optical control of Multiple Light Degrees of Freedom through Mode-mixing in a Graphene Nanoribbon Metamaterial

*Nikolaos Matthaiakakis,[1,*] Sotiris Droulias,[2,*] George Kakarantzas[3]*

[1] Institute of Nanoscience and Nanotechnology, National Center for Scientific Research "Demokritos", 15341 Ag. Paraskevi, Greece

[2] *Department of Digital Systems, University of Piraeus, Piraeus 18534, Greece*

[3] *Theoretical and Physical Chemistry Institute, National Hellenic Research Foundation, NHRF, 48 Vassileos Constantinou Ave., 11635 Athens, Greece*

**Abstract:** The evolution of optical technologies necessitates advanced solutions for selective and dynamic manipulation of light's degrees of freedom, including amplitude, phase, polarization, wavelength, and angular momentum. Metamaterials can offer such control through the interplay between the intrinsic material and geometrical properties of nanostructures or extrinsically through excitation and detection symmetry breaking, leading to customizable performance. However, achieving dynamic control over multiple light degrees of freedom remains a challenge. To address existing limitations, we present a novel dual-stack metamaterial design capable of broadband ultrafast control over amplitude, phase, polarization, spin angular momentum, and handedness of light mediated by two independently controlled nanoribbon layers that enable flexible and selective mode-mixing in both reflection and transmission. Through a combination of a thermal response model and Finite-Difference Time-Domain simulations, we investigate graphene as a suitable material for the metamaterial design, leveraging the intrinsic optical properties of graphene and its tunable conductivity through electrostatic gating and ultrafast optical excitation, achieving selective control over multiple light degrees of freedom at ultrafast timescales. This selective ultrafast mode-mixing significantly advances the capabilities of high-speed photonic systems, paving the way for compact, high data-rate optical technologies essential for future applications.

**Introduction**

The increasing demand for higher data rates, multifunctionality, and low latency in modern communication, logic, biomedical, environmental monitoring, and sensing systems, necessitates the development of advanced solutions to enhance manipulation capabilities over the light's degrees of freedom (LDOF) in ultrafast timescales, including the amplitude, phase, polarization, and angular momentum. For this purpose, the use of specially designed optical nanomaterials has grown significantly over the past decade, with a strong focus on two-dimensional (2D) structures that belong to the broader class of nanophotonics called metamaterials[1–5]. Novel metamaterials can enhance optical technologies by precisely controlling the LDOF and enabling highly directional and efficient light manipulation. For example, metamaterials can be used for polarization conversion and amplitude modulation[6–8]. Light manipulation with metamaterials is achieved through the exploitation of the properties that arise through the combined material properties and geometric design. This gives rise to optical modes that can be optimized in a controlled manner by adjusting the geometrical parameters of the metamaterial. Complex LDOF control such as polarization and angular momentum conversion is typically achieved through the combination of multiple different optical modes and by taking advantage of the interference arising from their amplitude/phase relations[9]. The simultaneous excitation of multiple modes, i.e. mode mixing, can be achieved either intrinsically, purely based on the metamaterial geometry and material properties, or extrinsically by externally breaking the symmetry of the system through excitation or detection symmetry. While intrinsic mode-mixing typically depends on complex geometries to achieve complex LDOF control, designs utilizing extrinsic mode mixing can be simpler, including even symmetric meta-atoms such as spheres[10]. Numerous publications taking advantage of extrinsic mode-mixing have demonstrated its efficiency in offering control over different LDOF [9,11–17]. Nevertheless, achieving control over multiple LDOF with a single metamaterial is challenging since different functionalities require different amplitude/phase relations between the interfering modes. Even though extrinsic mode-mixing, through broken excitation/detection symmetry, selectively combines different modes in metamaterials to create variable amplitude/phase combinations, there are still limitations to how much the properties of the supported modes can be tuned since the mode properties remain tied to the static material/geometrical properties of the metamaterial.

To a certain degree, this limitation can be overcome by introducing an element of dynamic tunability in the metamaterial designs [6,7,18–21]. This can be achieved through external stimuli such as an electric or magnetic field, a temperature change, or mechanical deformation[18,20,21]. All-optical ultrafast tuning through the metamaterial exposure to high power femtosecond laser pulses is considered an especially promising strategy as it enables extremely fast response times and eliminates the need for complex electrical or mechanical components while enabling seamless integration with existing photonic systems[13,22–29]. Nevertheless, several existing methods rely on changes of material conductivity, which in turn results in a global shift of all modes supported by the metamaterial. Hence, to achieve selective mixing of amplitude/phase relations between different modes, it is necessary that they can be independently controlled. Such a functionality would allow for unprecedented freedom over the control of multiple LDOF, offering high data rates and multiplexing capabilities which are crucial for future all-in-one optical and communication technologies such as 6G+ networks which aim for a unified network approach combining remote medicine, environmental monitoring, security, neuromorphic and quantum computing [7,30–33].

To tackle this challenge, we present a methodology for metamaterial design that combines the strengths of extrinsic mode mixing and all-optical ultrafast control of material conductivity, and the ability to selectively control individual modes in ultrafast timescales by utilizing a novel dual-stack meta-atom approach which separates the metamaterial modes into two independently controlled layers. The presented approach offers on demand flexible and efficient control over the mode-mixing both in reflection and transmission. The two layers of the metamaterial are two patterned graphene nanoribbon arrays, separated by a thin dielectric, and orthogonally rotated relevant to each other. The ability to control graphene's conductivity in various ways (chemical doping,[34] acoustic excitations[35], electrical gating.[36–40], optical excitation [41–43]), as well as its rapid thermal heating and cooling characteristics, make graphene an excellent material for providing the necessary tunability in the terahertz to mid-infrared spectral range[44,45]. We showcase the operating principles of the metamaterial through Finite-Difference Time-Domain (FDTD) simulations in combination with an experimentally verified model of the thermal response of graphene [27], demonstrating flexible mode-mixing, both electrostatically and in ultrafast timescales through all-optical pump-induced free carrier heating, achieving broadband control over the amplitude, phase, optical rotation, and spin angular momentum (including handedness control). To the best of our knowledge such flexibility in the control of multiple LDOF from a single metamaterial and especially in ultrafast timescales has never been demonstrated to this day in the terahertz – mid IR spectral range.

Through our findings we aim to provide a methodology for the realization of ultrafast control of multiple LDOF in ultrafast photonic systems, using compact and extremely thin nanophotonic devices.

**Proposed design and principle of operation**

The metamaterial consists of two arrays of single-layer graphene nanoribbons with a width (w) of 80nm and a periodicity (p) of 88nm, with each array being orthogonal to the other and separated by a 30nm thick dielectric with n=1.2 as seen in Figure 1a; thus, each layer is supporting localized surface plasmon (LSP) modes in orthogonal directions. The two layers are supported by a substrate of the same refractive index. The graphene conductivity is given by

$$\sigma(\omega) = \frac{D}{\pi(\Gamma - i\omega)}, \quad (1)$$

where $\omega$ is the radial frequency, $\Gamma$ is the scattering rate and $D$ is the Drude weight of graphene. The parameters $\Gamma$ and $D$ are temperature-dependent and, according to the experimentally verified thermal model[27], are given by

$$\Gamma(T) \simeq \Gamma_0 \left(1 + \frac{\pi^2 k_B^2 T^2}{6 E_F^2}\right) + \frac{E_F V_D^2 k_B T}{4\hbar^3 v_F^2 \rho s^2} \quad (2)$$

and

$$D(T) \simeq \frac{2e^2 E_F}{\hbar^2} \left(1 - \frac{\pi^2 k_B^2 T^2}{6 E_F^2}\right) \quad (3)$$

respectively, where $T$ is the temperature, $E_F$ is the Fermi energy, $v_f$ is the Fermi velocity, $e$ the electron charge, $\hbar$ is the reduced Planck constant, $k_B$ is the Boltzmann constant, $s$ is the speed of sound in graphene, $\rho$ is the areal mass density, $V_D$ is the acoustic deformation potential, and $\Gamma_0 = \frac{e v_f^2}{\mu E_F}$ [46].

The examples in this work have been optimized for graphene mobility of $\mu$=27000 cm$^2$ V$^{-1}$ s$^{-1}$, although much higher mobility values are experimentally possible[47–49] (see supplementary information for examples with other mobility values).

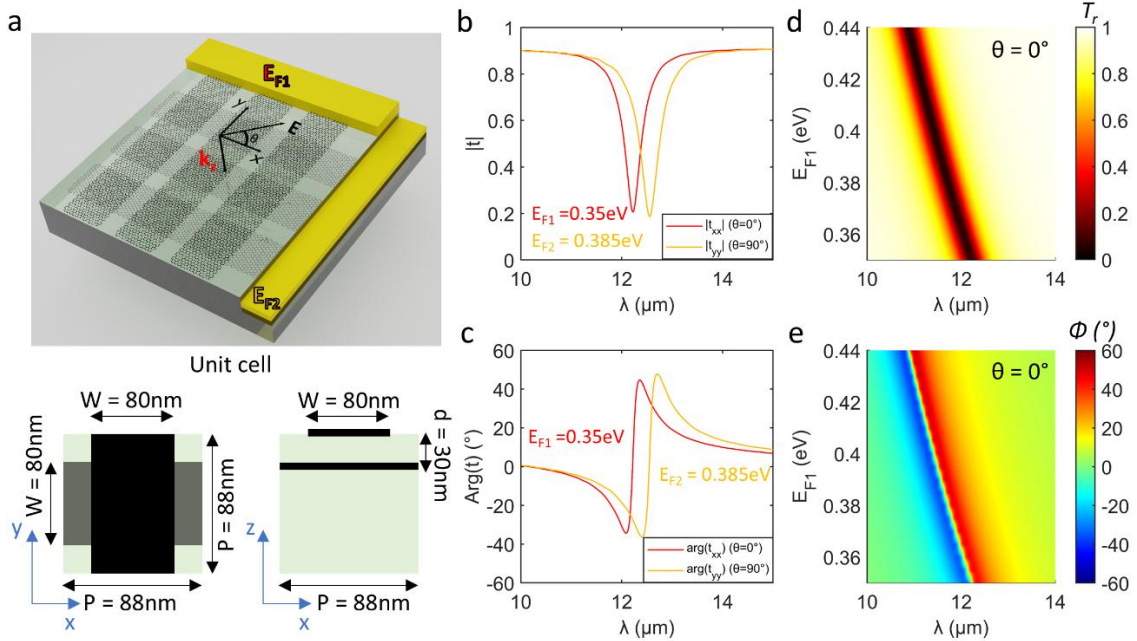

**Figure 1.** Graphene metamaterial for the control of multiple LDOF. a) Schematic of the dual stack graphene nanoribbon metamaterial (ribbon width 80nm, periodicity 88nm, ribbon 1 parallel to y direction, ribbon 2 parallel to x-direction and buried 30nm in the dielectric). b) Magnitude and c) phase of the transmission coefficients of the graphene for linearly polarized wave along the x- and y- direction (θ = 0° and θ = 90°, respectively). Broadband dynamic electrostatic control of d) transmittance $T_r$ and e) phase of transmitted light for θ = 0° for different values of $E_{F1}$ which corresponds to the Fermi level of ribbon array 1.

The metamaterial optical response is studied with the use of FDTD simulations (MEEP[50]). Details on the simulation process can be found in the supplementary information. First, we study the response of the metamaterial by probing the structure with a linearly polarized light of magnitude $E_0$. In the simulations, the azimuth angle (θ) of the probe source polarization is defined with respect to the x-axis, as illustrated in Figure 1a, which is analysed in the x- and y- directions as $E_{x,inc} = E_0 \cos\theta$ and $E_{y,inc} = E_0 \sin\theta$. The scattering matrix for the transmitted wave is written as:

$$\begin{pmatrix} E_{x,\text{tr}} \\ E_{y,\text{tr}} \end{pmatrix} = \begin{pmatrix} t_{xx} & t_{xy} \\ t_{yx} & t_{yy} \end{pmatrix} \begin{pmatrix} E_{x,\text{inc}} \\ E_{y,\text{inc}} \end{pmatrix} \quad (4)$$

where, $t_{xx}$, $t_{yy}$, $t_{xy}$, $t_{yx}$ are the transmission coefficients, $E_{tr}$ is the transmitted and $E_{inc}$ is the incident electric field, and the x,y subscripts denote the field components along the x- and y- directions, respectively. For incident polarization parallel to the x- or y- axis, we find that $t_{xy} = 0$, $t_{yx} = 0$ and $t_{xx} \neq 0$, $t_{yy} \neq 0$, thus the contributions from the x- and y-polarized components to the transmitted spectra under any given orientation of the incident light can be described through the transmission $t_{xx}$ and $t_{yy}$ coefficients, which are functions of $E_F$ and $T$, as

$$\mathbf{E}_{tr} = t_{xx}(E_{F1},T_1)E_{x,inc}\mathbf{x} + t_{yy}(E_{F2},T_2)E_{y,inc}\mathbf{y} \quad (5)$$

The top graphene ribbon array (ribbon array 1) is parallel to the *y* direction and can thus only support a radiative dipole mode in the *x*-direction perpendicular to the ribbons (LSP$_x$). Similarly, the buried ribbon array (ribbon array 2) which is instead parallel to the *x*-direction can only support a radiative plasmon mode in the *y*-direction (LSP$_y$), thus the metamaterial can support two LSPs that are orthogonal to each other, and each can only couple to either *x*- or *y*- polarized light, respectively. Figure 1.b shows the transmission amplitudes |$t_{xx}$| (|$t_{yy}$|) showing a peak reduction in the transmission corresponding to the LSP$_x$ (LSP$_y$) that is supported by ribbon array 1 (ribbon array 2). These values are controlled via the Fermi level of each ribbon array ($E_{F1}$ and $E_{F2}$) and electron temperature ($T_1$ and $T_2$). The corresponding phase for the transmission coefficients can be seen in Figure 1.c. The metamaterial response along the *x*- and *y*- direction can be independently controlled by selectively tuning $E_{F1}$ or $T_1$ to tune the response of the LSP$_x$ and $E_{F2}$ or $T_2$ to tune the response of LSP$_y$. In Figure 1.d,e we use an incident wave with θ = 0° (hence t$_{yy}$ = 0) to isolate the response of ribbon array 1, and we demonstrate how the transmittance $T_r$ = |$t_{xx}$|² and phase *Φ* = arg($t_{xx}$) of the transmitted light can be tuned across a wide spectral range through changes solely in $E_{F1}$. It is shown that as $E_{F1}$ increases there is a blue shift in the resonance wavelength of the LSP$_x$ resulting in a direct change in the transmission amplitude and phase for a given wavelength of choice and the effect can be controlled in a broadband manner achieving a shift of 1μm in wavelength for only a change of 0.09eV in $E_F$. For incidence with θ = 45° both LSP$_x$ and LSP$_y$ modes are excited, and their resonant frequencies can be tuned individually through $E_{F1}$ and $E_{F2}$, respectively, to spectrally shift the two modes together or independently. In this manner the amplitude and phase of the total transmitted field can be shaped through flexible mode mixing of the LSP$_x$ and LSP$_y$ response, allowing control over the amplitude and phase relations in any direction, as well as the ability to induce optical rotation or to introduce angular momentum since all these effects depend on the controlled superposition of amplitude and phase relations in the *x*- and *y*-directions. For example, linear to circular polarized conversion can be achieved when $t_{yy}(E_{F2},T_2)E_{y,inc} = t_{xx}(E_{F1},T_1)E_{x,inc}$ and $\arg(t_{yy}(E_{F2},T_2)) - \arg(t_{xx}(E_{F1},T_1)) = 90°$.

**Electrostatic control of multiple degrees of freedom of light**

Here we study our metasurface at room temperature, to demonstrate how dynamic control of multiple LDOF in transmission is possible through the selective tuning of the Fermi level ($E_{F1}$, $E_{F2}$) in the two nanoribbon arrays (see supplementary information for operation in reflection). Figure 2.a-c shows how different combinations of $E_{F1}$ and $E_{F2}$ can be used to tailor the amplitude and phase relations in the mode-interference, thus providing flexible control over multiple LDOF like the ellipticity $\eta$, and rotation $\alpha$ of the transmitted wave polarization, under linearly polarized light illumination with θ = 45°. In Figure 2.a-c the Fermi level of ribbon array 1 is kept at $E_{F1}$=0.35eV while $E_{F2}$ varies from 0.35eV to 0.5eV. This way, while the resonance wavelength of LSP$_x$ does not shift, LSP$_y$ can be shifted from longer to shorter wavelengths, as can be observed in the transmittance $T_r = |t_{xx}|^2+|t_{yy}|^2$ shown in Fig. 2.a. As LSP$_y$ is shifted the metamaterial enters different modes of operation. As seen in Figure 2.b, for $E_{F2}$ < 0.4eV the metamaterial operates as a linear-to elliptical/circular polarization converter with the ellipticity value increasing and approaching -45degrees as LSP$_x$ and LSP$_y$ shift closer together at $E_{F2}$=0.395eV and their amplitude/phase relations satisfy the requirements for circular polarization conversion (handedness convention from the point of view of the metamaterial). When $E_{F2}$=0.405 LSP$_x$ and LSP$_y$ spectrally overlap and the ellipticity is reduced to near zero and the metamaterial operates as an amplitude/phase modulator. For $E_{F2}$ > 0.4eV the metamaterial once again operates as a linear-elliptical/circular polarization converter with the opposite handedness when compared to the $E_{F2}$ < 0.4eV range. This handedness flip occurs due to the change of the sign of the phase difference between LSP$_x$ and LSP$_y$ as LSP$_y$ shifts from longer to shorter wavelengths with respect to LSP$_x$. Figure 2.c shows the modulation of rotation $\alpha$ for different $E_{F2}$ values. By comparing Figure 2.b and 2.c it is evident that significant changes in rotation $\alpha$ can occur for certain combinations of $E_{F1}$, $E_{F2}$, while $\eta$ approaches 0. It is thus possible to achieve controllable optical rotation of linearly polarized light.

Thus far we have demonstrated that, by independently controlling $E_{F1}$ and $E_{F2}$, the metamaterial can provide selective control over the amplitude, phase, ellipticity (including handedness), and optical rotation. It is also useful to extend this control in the wavelength domain. This can be achieved by first choosing the desired LDOF control by selecting a combination of $E_{F1}$ and $E_{F2}$ based on the results of Figure 2.a-c. For this example, we select $E_{F2}$=$E_{F1}$+0.035eV which sets the metamaterial to operate as a linear to circular polarization converter, as shown in Figure 2.d-f. By increasing or decreasing $E_{F2}$ and $E_{F1}$ equally, LSP$_x$ and LSP$_y$ shift together (see Fig. 2.d) and the amplitude/phase relations between the two modes remain unchanged. The metamaterial can thus operate as a linear to circular polarization converter over a broad spectral range (Figure 2.e). In this configuration. the metamaterial can also operate as an optical rotator in a broad spectral range as seen in Figure 2.f.

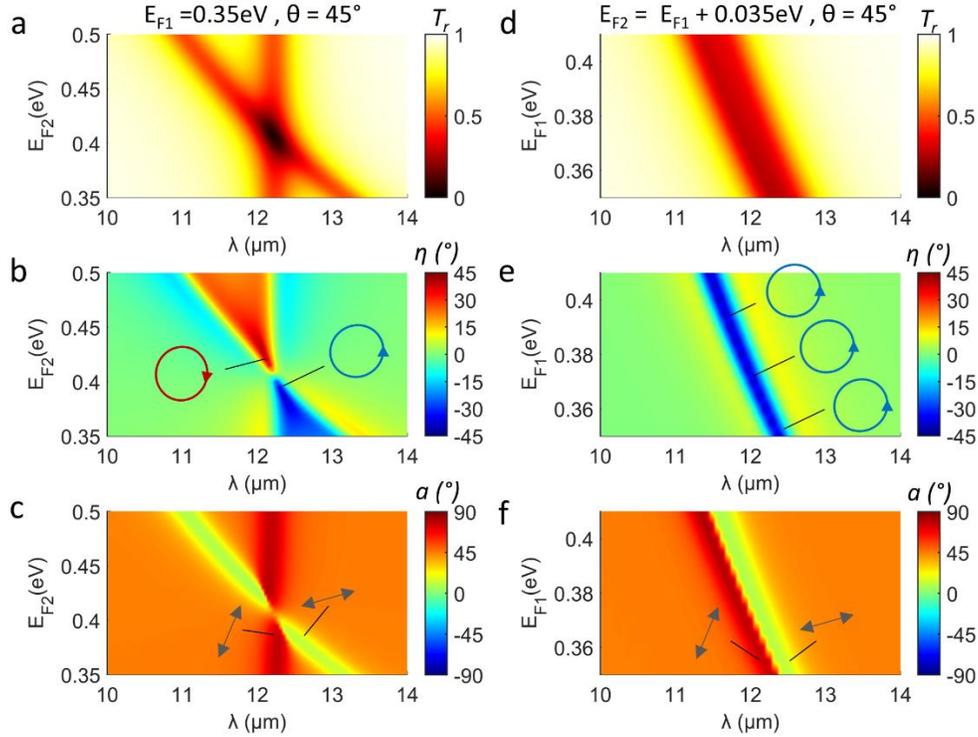

**Figure 2.** Utilization of flexible mode-mixing through independent dynamic control of the x- and y-polarized plasmon modes supported by the two orthogonal graphene nanoribbons to achieve broadband dynamic control over the amplitude, ellipticity, and rotation angle of the transmitted light. Dynamic electrostatic control over the a) transmittance, b) ellipticity $\eta$ and c) rotation $\alpha$ for transmitted light for $E_{F1}$=0.35eV while $E_{F2}$ varies from 0.35eV to 0.5eV. In g-i) the combination of $E_{F1}$ and $E_{F2}$ that achieves maximum ellipticity in (b) is selected and $E_{F1}$ and $E_{F2}$ are tuned together to achieve broadband tunability of the linear to circular polarized conversion. Broadband tuning of the d) transmittance, e) ellipticity, f) rotation for transmitted light.

By selecting different combinations of $E_{F2}$ and $E_{F1}$, broadband control over all the types of LDOF control supported by the metamaterial is possible. It is thus useful to map the response of the metamaterial for different combinations of $E_{F2}$ and $E_{F1}$. Figure 3.a-c shows the transmittance, ellipticity $\eta$, and rotation $\alpha$ through the metamaterial at λ=12.3μm for a range of $E_{F1}$ between 0.3eV and 0.4eV and $E_{F2}$ from 0.35eV to 0.45eV. By independently controlling LSP$_x$ and LSP$_y$ a variety of different operational states for the metamaterial is available for selection. Most importantly, it is clear form Figure 3.a-c that different combinations of $E_{F1}$ and $E_{F2}$, can result in very different combinations of $T_r$, $\eta$, and $\alpha$, allowing us to choose clearly defined states of operation such as, amplitude modulation of linearly polarized light (regions where $\eta=0°$ and variable $T_r$) or elliptically polarized light (regions where $\eta\neq0°$ and variable $T_r$), optical rotation of linearly polarized light (regions of $\eta=0°$ and variable $\alpha$), ellipticity modulation (regions of variable $\eta$), linearly to circular polarization conversion of either handedness (regions where $\eta \approx \pm 45°$). Figure 3.d demonstrates that the desired metamaterial response, in this case linear-to-circular polarization conversion for both left and right handedness, can be tuned in the wavelength domain if the appropriate combination of $E_{F1}$ and $E_{F2}$ is used. In this case, as the values of $E_{F1}$ and $E_{F2}$ increase, there is a linear relationship between the two values of Fermi level that can be followed to shift the response in the wavelength domain. Thus, through the combination of $E_{F1}$ and $E_{F2}$ the metamaterial can provide broadband flexible control over the amplitude, rotation, ellipticity, and handedness of the transmitted light.

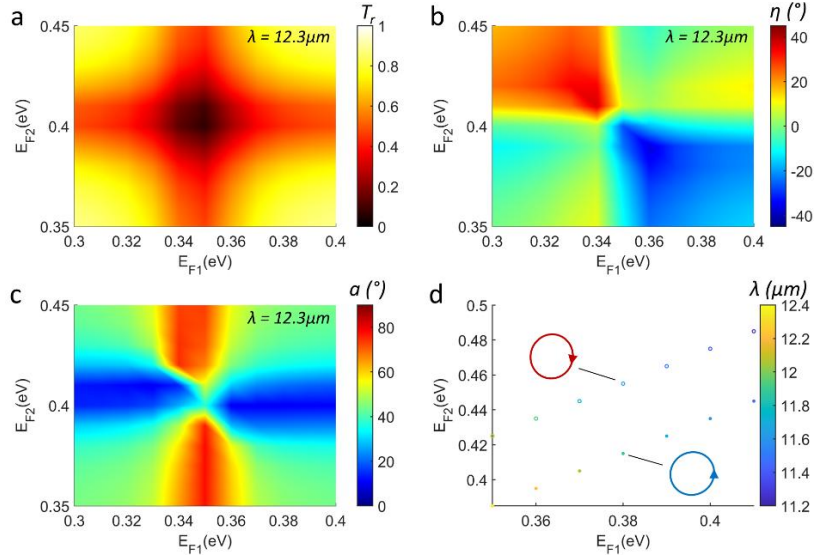

**Figure 3.** Dynamic control of a) transmittance, b) ellipticity $\eta$, c) rotation $\alpha$ for different combinations of $E_{F1}$ and $E_{F2}$ at $\lambda=12.3\mu m$. d) Broadband tunability of linear-to-circular polarization conversion, as a function of $E_{F1}$ and $E_{F2}$; the solid (open) circles correspond to LCP (RCP) waves.

**All-optical ultrafast control of multiple degrees of freedom of light**

In this section we demonstrate how the electrostatic tuning of the metamaterial can be extended by introducing pump induced all-optical ultrafast control to take advantage of selective electron thermalisation in the two graphene nanoribbon array layers. The effect of electron thermalization is included in the model through $T$ in equations 1-3. In practice such electron thermalization can be achieved by introducing a pump pulse in resonance with the plasmon resonance frequency supported by the metasurface. The electron temperature $T$ in graphene under pump pulse excitation with intensity I(t) and a central frequency $\omega_0$ evolves as

$$\alpha T \frac{dT}{dt} + \beta(T^3 - T_L^3) = A(\omega_0; T)I(t) \quad (7)$$

where $\alpha T$ is the specific heat of graphene with $\alpha = 2\pi k_B^2 E_F/(3\hbar^2 v_F^2)$, $\beta = \zeta(3)V_D^2 E_F k_B^3 /(\pi^2 \rho \hbar^4 v_F^3 s^2 l)$ is the cooling coefficient, $A(\omega)$ is the fractional absorption in the graphene, $\zeta$ is the Riemann zeta function, $l$ is the electron-disorder mean free path, and $T_L$ is the lattice temperature [27]. For the narrow-band terahertz - mid IR pulses analysed in this study, we can estimate the relationship between the peak temperature ($T$) and fluence ($F$) by assuming that the peak temperature and peak intensity occur nearly simultaneously in time, that the pump pulse frequency coincides with the frequency of the maximum absorption max($A(\omega)$), and neglecting the lattice temperature and higher-order temperature dependent changes in the absorption coefficient $A$ as [27]

$$\beta T_{max}^3 = A(\omega)I_{max} \quad (8)$$

where $T_{max}$ is the peak electron temperature and $I_{max}=0.94F/\Delta t_{FWHM}$ is the peak intensity of a Gaussian pulse with fluence $F$ and duration $\Delta t_{FWHM}$. Equation 7 is solved with the use of the Euler method, and equations 1,2, and 3, for the Drude weight, scattering losses, and conductivity, are updated at each time step of the solution of equation 7 according to the calculated electron thermalization. The absorption $A(\omega_0, T)$ for different values of graphene electron temperature $T$ is obtained through FDTD simulations and is adjusted accordingly at each individual time step of the solution of equation 7.

Figure 4a shows how a pump pulse can be used to control $LSP_x$ in all-optical and ultrafast manner for a probe polarization of $\theta_{pr} = 0°$. To achieve selective tuning of $LSP_x$ a pump polarization $\theta_{pu} = 0°$ is selected, thus being perpendicular to ribbon array 1 and parallel to ribbon array 2. The pump pulse is quasi-monochromatic (5.5ps duration) and is launched at a wavelength of 13µm to spectrally overlap with the tail of the $LSP_x$ ($E_{F1}$=0.35eV, see Fig.1b). The chosen pump polarization guarantees that only $LSP_x$ is excited, and the pump beam is mainly absorbed by $LSP_x$ in ribbon array 1, as $LSP_y$ in ribbon array 2 cannot couple to x-polarized light. This results in elevated temperature in only ribbon array 1 and in a redshift and broadening of $LSP_x$ due to the heightened scattering rates and Fermi level shift from the carrier thermalization. The pump pulse has fluence F=2.22 µJ/cm², corresponding to energy levels similar to the experimental work of [27], well below the damage threshold of graphene[51–53]. Complete electron relaxation process throughout this work takes place in good agreement with previous theoretical and experimental works[27,54,55]. The all-optical tuning of the transmission amplitude and phase through the metamaterial due to the pump induced electron thermalisation can be seen in Figure 4.b,c respectively. The metamaterial demonstrates broadband shift of the transmission peak, as well as significant control over the phase of the transmitted light in ultrafast timescales, offering an all-optical equivalent to the electrostatically tunable amplitude and phase modulator of Figure 1.d,e (see supplementary information for operation in reflection).

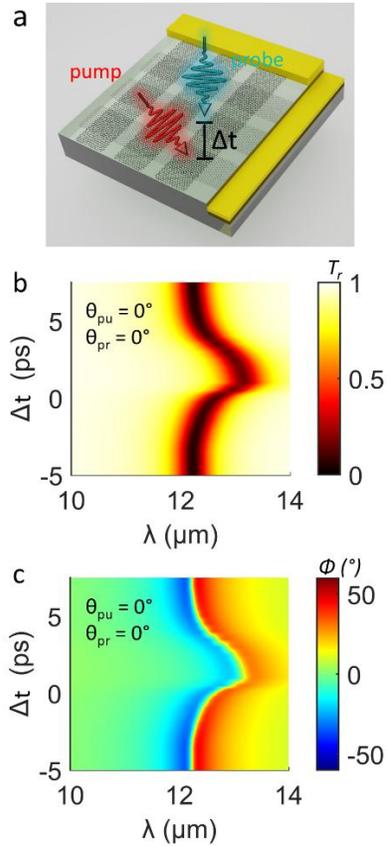

**Figure 4.** All-optical ultrafast operation of the graphene nanoribbon array metamaterial. a) Schematic showing the interaction of the metamaterial with the pump and the probe pulses. Broadband all-optical ultrafast control of b) transmittance and c) phase of transmitted light. The results are obtained for $\theta_{pu}$ =0°, $\theta_{pr}$ =0° and $E_{F1}$=0.35eV, $E_{F2}$=0.385eV.

Selective all-optical pumping including both ribbon layer arrays can also be used to achieve control over multiple LDOF in an ultrafast all-optical manner. Any combination of $E_{F1}$ and $E_{F2}$ of Figure 3 can be chosen upon which all-optical tuning can be applied to. For Figure 5a-f $E_{F1}$=0.35eV and $E_{F2}$=0.385eV are selected, thus at rest (i.e. without pump excitation) the metamaterial operates as a linear-to-circular polarization converter. For Figure 5.a-c the pump polarization rotation $\theta_{pu} = 0°$ and probe polarization rotation $\theta_{pr} = 45°$, thus the pump pulse interacts mainly with $LSP_x$, resulting in selective electron thermalization of ribbon array 1. When at rest, before the interaction of the pump pulse with ribbon array 1, $LSP_x$ is slightly blue-shifted compared to $LSP_y$ and $\eta$ approaches -45° in transmission. As the electron temperature in ribbon array 1 increases, $LSP_x$ is redshifted (Figure 5a), moving to the same wavelength with $LSP_y$ at $\Delta t=0$ps, and then crossing over $LSP_y$ to longer wavelengths for $0<\Delta t<4$ps, finally returning to its original position as the electron temperature is cooling down for $\Delta t>2.5$ps, once again crossing over $LSP_y$ at $\Delta t=4$ps. As $LSP_x$ shifts closer or further apart from $LSP_y$ the amplitude and phase relations in the x- and y-directions are significantly altered resulting in changes in the ellipticity $\eta$ and rotation $\alpha$ (Figure 5.b,c). Initially, for $\Delta t<0$ps while $LSP_x$ is at shorter wavelengths compared to $LSP_y$, looking at the wavelength of maximum $\eta$, as the electron thermalization increases $\eta$ decreases until it reaches $\eta = 0°$ at $\Delta t=0$ps when $LSP_x$ and $LSP_y$ are at the same wavelength. For $\Delta t>0$ps when $LSP_x$ has crossed over $LSP_y$ to longer wavelengths the phase difference between the modes swaps sign and thus there is a handedness switch for $\eta$, achieving linear to elliptical or circular polarization of the opposite handedness when compared to the operation at rest. As the electron temperature begins to cool down $\Delta t>2.5$ps, $\eta$ reverts to its original state. The biggest changes in $\alpha$ occur at different wavelengths compared to $\eta$ and especially at regions where $\eta$ approaches 0°. As a result, the selectivity in the all-optical dynamic control of $\eta$ and $\alpha$ is maintained as was the case for the electro-optical control of Figure 3 and all-optical ultrafast control of the rotation $\alpha$ linearly polarized light is also possible (Figure 5.c).

It is also possible to use the pump pulse to control the electron temperature in both ribbon array 1 and ribbon array 2 at the same time to obtain broadband ultrafast control of the device properties in the wavelength domain. This can be achieved by setting $\theta_{pu} = 45°$ to equally couple the pump beam with $LSP_x$ and $LSP_y$ and probe the metasurface with $\theta_{pr} = 45°$. This way the pump can tune the resonant wavelength of both LSPs at the same time, thus maintaining the phase and amplitude relations in the x- and y- direction while shifting the response in the wavelength domain. This is demonstrated in Figure 5.d-f, where a pump pulse at $\lambda=13$μm with $\theta_{pu} = 45°$ (thus interacting with both $LSP_x$ and $LSP_y$), considering F=2.22 μJ/cm$^2$ in both the x- and y- directions, is used to tune the two modes together. From Figure 5.d-f we can see that the two LSPs are red-shifting together as the electron temperature increases. Nevertheless, as $E_{F1}$ and $E_{F2}$ have different values and since the temperature dependent Drude weight and scattering rates depend on $E_F$, there is a small difference in the rate that $LSP_x$ and $LSP_y$ shift. An additional parameter that contributes to this issue is that as LSP$_x$ and LSP$_y$ shift, their interaction with the pump pulse through the absorption A (see equation 7) can also result in an unequal electron thermalization for ribbon array 1 and ribbon array 2, since $A_x \neq A_y$ for the same λ. As a result, there is a fluctuation in the amplitude/phase relations of the two modes as they are shifted in frequency. This is more obvious in Figure 5.e since η is very sensitive to the amplitude and phase relations between LSP$_x$ and LSP$_y$ and thus a narrowing and weakening of the η peak can be observed. Such slight amplitude/phase detuning can be taken into account, to tailor the desired performance and design a metamaterial that can be tuned all-optically at ultrafast timescales, across a wide spectral range (see supplementary information for operation in reflection).

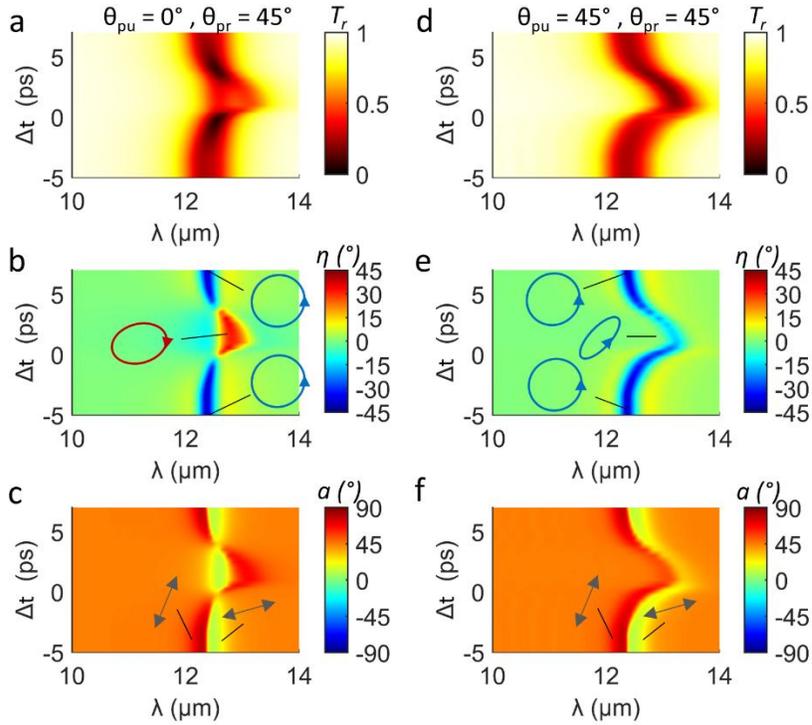

**Figure 5.** All-optical ultrafast independent control of the x- and y-polarized plasmon modes supported by the two orthogonal graphene nanoribbons. a) transmittance, b) ellipticity η and c) rotation α of transmitted light for $E_{F1}$=0.35eV, $E_{F2}$=0.385eV, and θ$_{pu}$ =0°, θ$_{pr}$ =45°. Ultrafast broadband tuning of d) transmittance, e) ellipticity η and f) rotation α of transmitted light for θ$_{pu}$ =45°, θ$_{pr}$ =45°.

**Discussion and conclusions**

In conclusion, this work presents a novel metamaterial platform capable of ultrafast, selective control over multiple LDOF. By integrating extrinsic mode mixing with all-optical ultrafast control of material conductivity, split into two independently controlled layers of graphene nanoribbon arrays, the proposed metasurface design offers unprecedented flexibility in amplitude, phase, polarization, and spin angular momentum manipulation. Our results, supported by FDTD simulations and an experimentally validated graphene thermal response model, demonstrate that independent control over individual metamaterial modes can be realized on ultrafast timescales by leveraging graphene's intrinsic nonlinear optical properties and rapid thermal response, offering a scalable and integrable solution for ultrafast photonic systems. Importantly, the proposed metasurface offers a versatile platform for LDOF control both in transmission and reflection. This potentially allows for dynamic, real-time tunability, essential for future low latency and high data rate transmission networks. This capability is crucial for next-generation optical technologies, particularly in high-speed communication, neuromorphic computing, biomedical imaging, environmental monitoring, and security applications. Despite potential fabrication challenges, such as achieving high-mobility graphene and minimizing structural defects, our findings highlight the feasibility of practical implementations. By addressing key challenges in ultrafast optical control, our proposed design could open new paths in the development of next-generation metamaterials that enable high-speed, low-latency optical processing.

**Author information**

Corresponding Authors N. Matthaiakakis[1,*] , S. Droulias[2,*]

E-mail: n.mattheakakis@inn.demokritos.gr, sdroulias@unipi.gr

**Author Contributions**

The manuscript was written through contributions of all authors. NM conceived the idea of this work and performed the simulations and theoretical calculations. SD and GK provided supervision of the project. All authors have given approval to the final version of the manuscript.

**Disclosures**

The authors declare no conflicts of interest.